# Approche pédagogique sur l'innocuité des technologies de réseaux sans fil

François Demontoux, Rafael Hidalgo Muñoz
Université Bordeaux, IUT GEII Talence et Laboratoire IMS, 16. av. Pey Berland 33607 Pessac,
francois.demontoux@ims-bordeaux.fr

**RESUME :** Les technologies de réseaux sans fil sont de plus en plus répandues. Dans le cadre de nos enseignements, nous sommes amenés à aborder les aspects techniques de ces technologies. Toutefois, l'expérience de la téléphonie mobile nous a montré les craintes que peuvent susciter ces produits auprès du grand public. Ces dernières proviennent de l'utilisation d'ondes électromagnétiques à des fréquences proches des fréquences de nos fours micro-ondes ménagers (2.45 GHz). Des études sont en cours depuis des années afin de déceler des effets éventuels sur l'organisme liés à l'utilisation prolongée des téléphones portables. Les réseaux sans fil de type Wifi ou ZigBee par exemple utilisent aussi des fréquences micro-ondes. D'autre part, les problèmes liés à la compatibilité électromagnétique des ondes avec les appareils électroniques doivent aussi être étudiés. La question de l'innocuité de ces technologies doit donc être abordée avec les étudiants. Ceux-ci pourront être amenés à installer ce type de matériel. Nous avons intégré à l'enseignement des réseaux sans fil un chapitre de cours traitant des normes d'exposition aux ondes électromagnétiques pour les êtres vivants et les appareils électroniques. Le but de notre démarche est de faire acquérir par l'étudiant des connaissances afin qu'il se forge une opinion et puisse fournir un discours éclairé aux futurs utilisateurs de ces réseaux. A cette fin, nous avons mis au point un outil de visualisation des répartitions de champs électromagnétiques. Cet outil utilisé en cours permet de comprendre l'environnement électromagnétique dans lequel nous vivons et de quantifier les valeurs de champs électromagnétiques qui nous traversent en fonction des applications (téléphonie, télévision, radio, WiFi, Bluetooth, ZigBee, four micro ondes …).

**Mots clés** : Interactions ondes électromagnétiques - être humain et appareils électronique, compatibilité électromagnétique, innocuité des applications de réseaux sans fil, WiFi, ZigBee, simulation numérique, propagation.

## 1 INTRODUCTION

Les technologies de réseaux sans fil (WiFi, ZigBee, Wimax …) sont de plus en plus répandues. Dans le cadre de nos enseignements, nous sommes amenés à aborder les aspects techniques de ces applications. Toutefois, l'expérience de la téléphonie mobile nous a montré les craintes que peuvent susciter ces produits auprès du grand public. Ces dernières proviennent de l'utilisation d'ondes électromagnétiques à des fréquences proches des fréquences de nos fours micro-ondes ménagers (2,45 GHz). Des études sont en cours depuis des années afin de déceler des effets éventuels sur l'organisme liés à l'utilisation prolongée de téléphone portable. Hors, les réseaux sans fil de type Wifi ou ZigBee par exemple utilisent aussi des fréquences micro-ondes (au voisinage de 2,4 GHz). Les problèmes liés à la compatibilité électromagnétique des ondes avec les appareils électroniques ou les êtres vivants doivent donc aussi être étudiés dans le cadre de l'utilisation de ces technologies.

Les réseaux sans fil de type Wifi ou ZigBee par exemple utilisent aussi des fréquences micro-ondes (au voisinage de 2,4 GHz). D'autre part, les problèmes liés à la compatibilité électromagnétique des ondes avec les appareils électroniques doivent aussi être étudiés.

La problématique de l'innocuité de ces technologies doit donc être abordée avec les étudiants. Ceux-ci pourront être amenés à installer ce type de matériel. Dans le cadre de la licence professionnelle SARI de l'IUT GEII de Talence (IUTA- Université Bordeaux 1), nous avons intégré à l'enseignement des réseaux sans fil un chapitre de cours traitant des normes d'exposition aux ondes électromagnétiques [6] [7] [8] [9] [10] pour les êtres vivants et les appareils électroniques. Le but de notre démarche est de faire acquérir par l'étudiant des connaissances afin qu'il se forge une opinion et puisse fournir un discours éclairé aux futurs utilisateurs de ces réseaux.

Le but du travail que nous présentons a été de mettre au point un outil de visualisation des répartitions de champs électromagnétiques. Cet outil, couplé avec des mesures effectuées au laboratoire IMS (Intégration du Matériau au Système - département MCM) permet de comprendre l'environnement électromagnétique dans lequel nous vivons et de quantifier les valeurs de champs électromagnétiques qui nous traversent en fonction des applications (téléphonie , télévision, radio, WiFi, Bluetooth, ZigBee, four micro ondes …). Cet outil est déjà utilisé en licence EISI (Electronique et Informatique des Systèmes Industriels) option SARI (Systèmes Automatisés – Réseaux Industriels).

## 2 PRESENTATION DES LOGICIELS DE SIMULATION

Le but des simulations que nous avons effectuées était de représenter la répartition des champs électromagnétiques des principales applications précédemment évoquées dans une pièce de 35,2m$^3$ (4m x 4m x 2,2m).

Le premier logiciel que nous avons utilisé est le logiciel HFSS (High Frequency Struture Simulator) de la société ANSOFT [2]. L'utilisation de cet outil dans le cadre de recherches scientifiques au sein du laboratoire IMS (Intégration du matériau au système) nous a permis d'obtenir de la société ANSOFT des licences gratuites pour l'enseignement.

Ce logiciel est basé sur la méthode des éléments finis [3]. C'est une méthode de résolution des équations de Maxwell qui nécessite le découpage de l'espace en tétraèdres sur lesquels s'applique l'algorithme de résolution. De par sa méthode de résolution numérique, ce logiciel est d'autant plus « gourmand » en mémoire RAM que la fréquence des ondes simulées augmente, c'est-à-dire que leur longueur d'onde diminue. Par exemple, à 3GHz, la longueur d'onde dans l'air est de 10cm. Si nous utilisons 10 tétraèdres par longueur d'ondes, les 35,2m$^3$ de la pièce seront « maillés » à l'aide de plus de 11.7 10$^6$ tétraèdres ce qui conduirait à l'utilisation d'une mémoire d'au moins 10Go …

Les contraintes de temps de calcul et de mémoire nous obligent à nous limiter au maximum à 60 000 cellules. La solution retenue a été de limiter notre problème à un problème pratiquement à 2 dimensions en limitant la hauteur de la pièce. Cette simplification du problème impose aux objets d'être représentés comme ayant la même hauteur que la pièce. Elle simplifie aussi les phénomènes de propagation. Nous verrons toutefois que ce modèle, déjà utilisé dans le cadre de travaux pratiques sur les réseaux WiFi [1] permet une très bonne compréhension des phénomènes de propagation.

La figure 1 représente la géométrie que nous avons simulée. Le volume de la pièce est de 1,6m$^3$ (4m x 4m x 10 cm). Nous y avons placé des cloisons et des armoires.

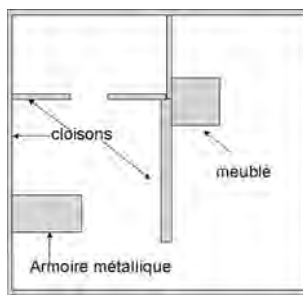

*Figure 1 : géométrie représentée - modèle HFSS*

Afin d'obtenir des résultats plus représentatifs de la réalité, nous avons développé à l'aide du logiciel Microstripes [4] (de la société Flomerics) un nouveau modèle 3D de notre problème. En effet la méthode de résolution utilisée (TLM : Transmission Line Matrix) nécessite beaucoup moins de ressources informatiques que le logiciel HFSS.

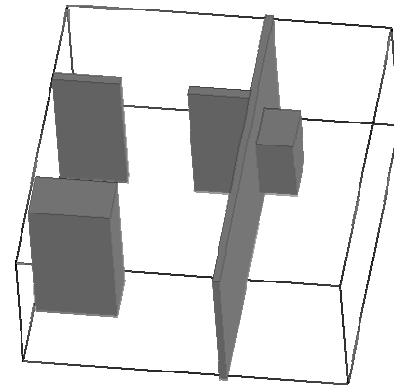

*Figure 2 : géométrie représentée – modèle 3D microstripes*

Les sources de génération des champs électromagnétiques seront représentées à l'aide de dipôles rayonnants. La position des différentes sources est indiquée sur la figure 3.

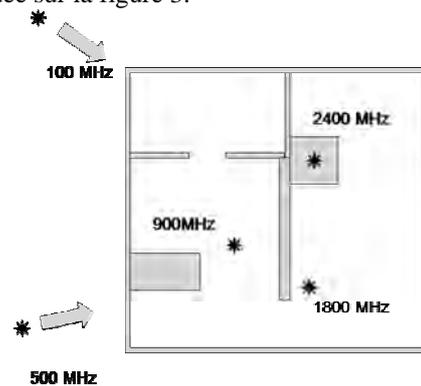

*Figure 3 : position des différentes sources*

## 3 SIMULATION DE LA PROPAGATION DU CHAMP ELECTROMAGNETIQUE A L'AIDE DU MODELE HFSS.

### 3.1 Simulation de la propagation du champ électromagnétique émis par un point d'accès WiFi (2,4 GHz)

L'ensemble des résultats de cartographie de champs électriques présentés est référencé à l'aide d'une échelle de couleur qui permet de repérer les niveaux maximums et minimums. Par la suite, ces cartographies seront normalisées afin de tenir compte des niveaux réels de puissance rencontrés pour les différentes applications. La pièce n'est pas simulée entièrement, la hauteur a en effet était limitée. Cette approximation simplifie les répartitions des champs électromagnétiques en limitant leur propagation dans un espace pratiquement à deux dimensions.

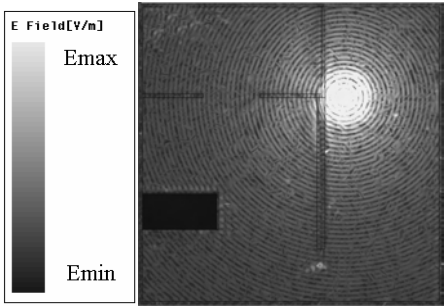

*Figure 4 : Application Wifi*

### 3.2 Simulation de la propagation du champ électromagnétique émis par des téléphones portables (900 et 1800 MHz)

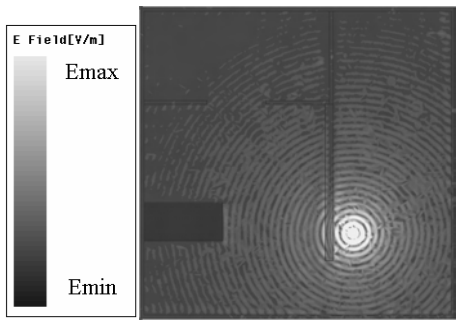

*Figure 5 : Application "téléphonie mobile 1800 MHz"*

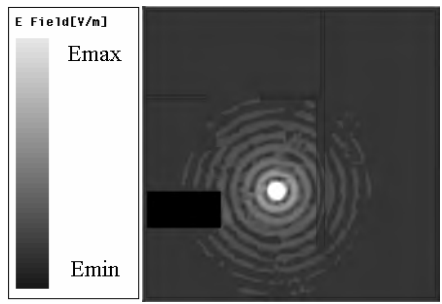

*Figure 6 : Application ˝téléphonie mobile 900MHz˝*

### 3.3 Simulation de la propagation du champ électromagnétique émis par la télévision (500 MHz)

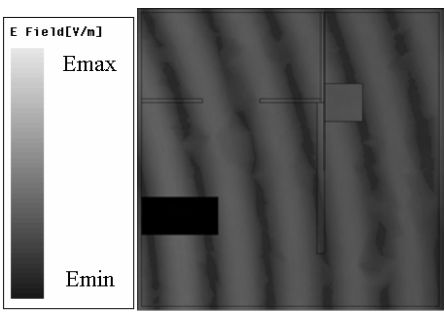

*Figure 7 : Application ˝télévision 500 MHz˝*

### 3.4 Simulation de la propagation du champ électromagnétique émis par la radio (100 MHz)

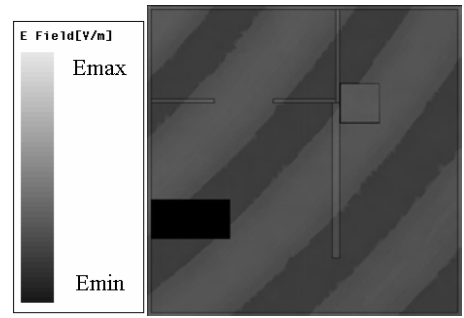

*Figure 8 : Application ˝radio 100 MHz˝*

## 4 RESULTATS DU MODELE 3D MICROSTRIPES .

Nous avons développé un modèle en trois dimensions réalisé avec le logiciel Microstripes.

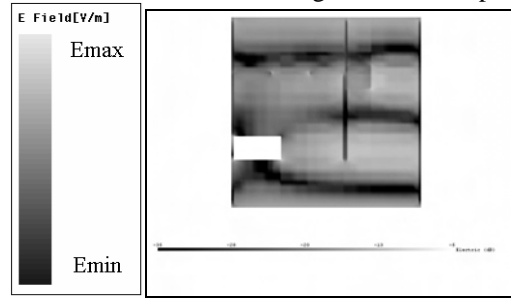

*Figure 9 : Application radio 100 MHz – coupe 2D*

Les résultats obtenus (figure 9 à 13) permettent de réaliser les principales différences de propagation de l'onde liées à la longueur d'onde de l'application. De plus, contrairement aux simulations précédentes, la prise en compte du volume total de la pièce provoque des phénomènes de propagations dans toutes les directions ce qui rompt la symétrie de répartition du champ électromagnétique observé précédemment.

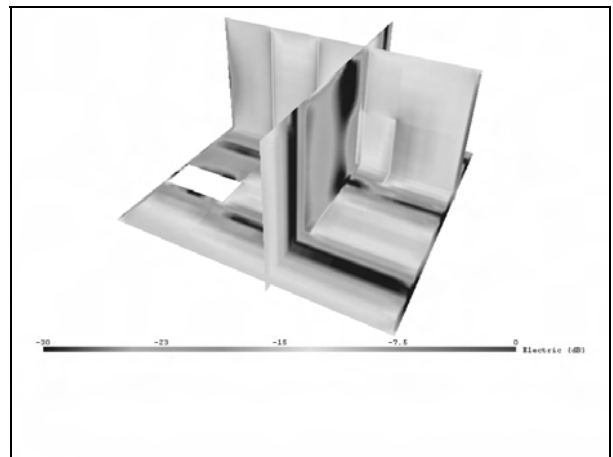

*Figure 10 : Application ˝ radio 100 MHz˝*

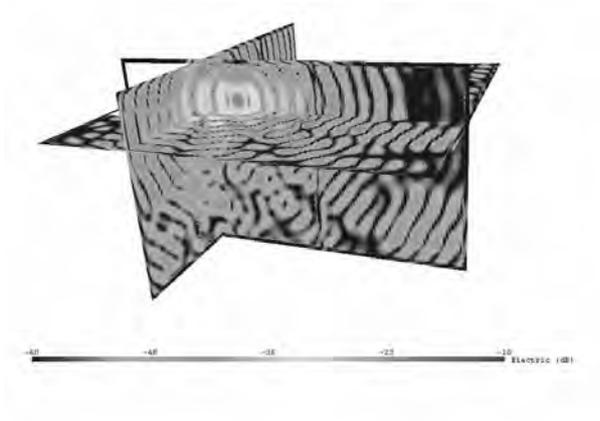

*Figure 11 : Application ˝ téléphone portable 900 MHz˝*

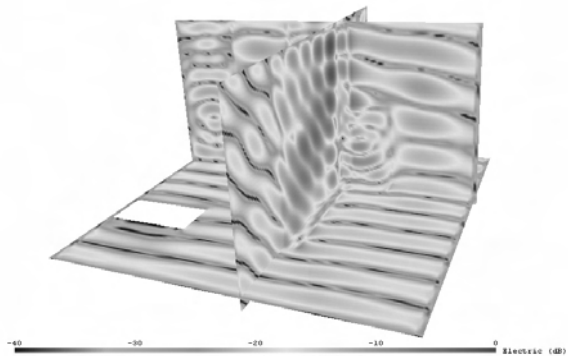

*Figure 12 : Application ˝ télévision 500 MHz˝*

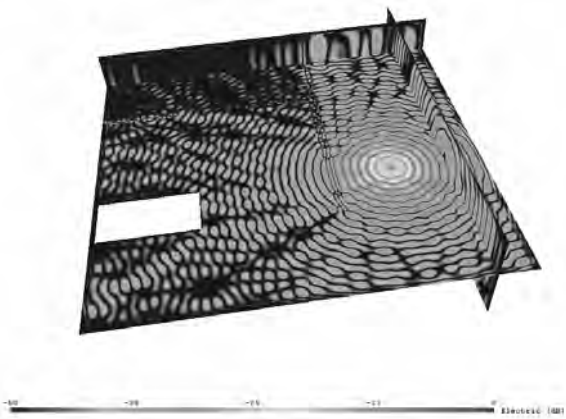

*Figure 13 : Application ˝ téléphonie 1800 MHz˝*

### 4.1 Simulation de la propagation du champ électromagnétique émis par une borne WiFi (2450 MHz)

La faible longueur d'onde de cette application nous a obligé à limiter l'espace de travail (figure 14). Toutefois, la faible puissance de la source limite d'elle-même la zone dans laquelle les effets de cette source peuvent interagir avec un organisme ou un appareil électronique.

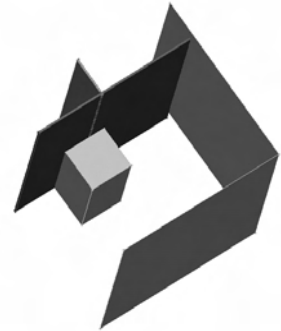

*Figure 14: Partie de la pièce simulée*

*(2m x 2 m x 2.2m)*

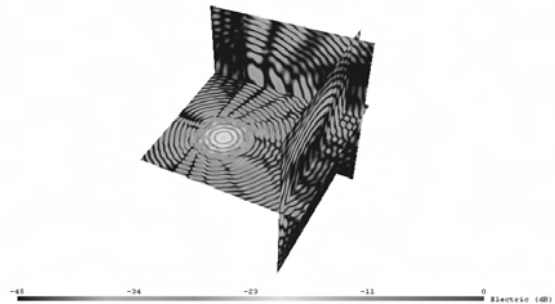

*Figure 15 : Application Wifi*

### 4.2 Simulation d'un utilisateur de téléphone portable (900 MHz)

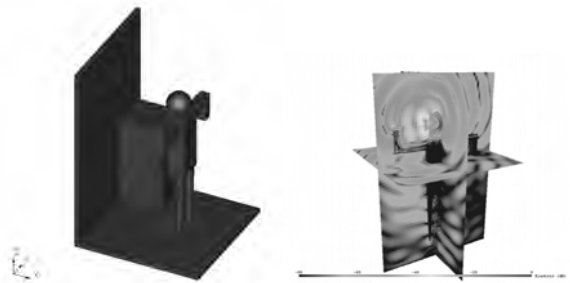

*Figure 16 : utilisateur de téléphonie mobile*

Cet exemple nous permet d'aborder les changements de longueurs d'ondes dans les milieux (ici un être humain), les notions de profondeur de pénétration ou de DAS (débit d'absorption spécifique). Des permittivités des principaux tissus humains sont intégrés dans le modèle [11].

### 5 MESURES DES NIVEAUX DE CHAMPS

Pour connaître les différents niveaux de puissance des sources, nous avons réalisé des expériences avec un analyseur de spectres.

Nous avons utilisé deux antennes adaptées à deux bandes de fréquences différentes; une antenne râteau pour la réception des plus basses fréquences (radio, télévision et téléphone portable à 900 MHz) et une antenne cornet plus directionnelle et plus sensible pour détecter les autres fréquences (téléphone portable à 1800 MHz et WiFi).

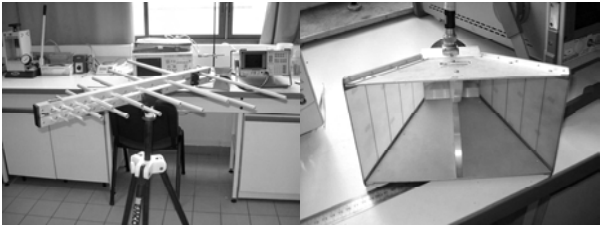

*Figure 17 : antennes utilisées pour les mesures*

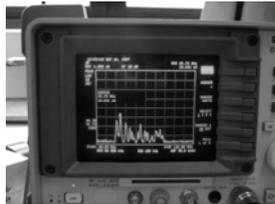

Radio (bande 80 MHz – 120 MHz).P=4nW

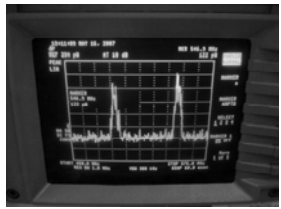

Télévision (bande 450 MHz – 500 MHz). P=140pW

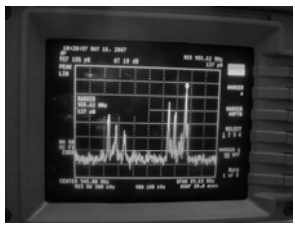

Téléphone portable (900 MHz). P=14nW

*Figure 18 : Mesures de puissances rayonnées*

Nous avons pris quelques mesures sur chaque bande de fréquences des différentes applications et nous avons calculé la valeur moyenne. Nous avons utilisé ces données pour la simulation et la création de l'échelle lors de la représentation des résultats.

Les mesures ont montré que l'application « télévision » présente le niveau de champ le plus faible, et que les champs provoqués par les téléphones portables sont toujours les plus importants. Il est important de noter que la puissance des ondes électromagnétiques dépend beaucoup de la distance à la source. Par exemple dans le cas du WiFi le niveau de champ est considérablement plus fort si le point d'accès WiFi est dans la pièce plutôt qu'à quelques mètres derrière une cloison (figure 4). Le tableau 1 résume les principaux résultats de nos mesures.

| Source | Fréquence utilisée | Longueur d'onde | Puissance mesurée |
|---|---|---|---|
| Radio | 100 MHz | 3 m | 4 nW |
| Télévision | 500 MHz | 60 cm | 140 pW |
| Portable | 900 MHz | 33 cm | 14 nW |
| Portable | 1800 MHz | 17 cm | 10 nW |
| WiFi | 2.4 GHz | 12.5 cm | 600 pW |

Tableau 1

## 6 NORMALISATION DES NIVEAUX DE PUISSANCE EMISE PAR LES DIFFERENTES APPLICATIONS

Notre objectif est de représenter l'environnement électromagnétique dans lequel nous vivons. Pour cela les niveaux de puissances émises par les différentes applications envisagées doivent être normalisés afin de tenir compte des niveaux réels. Les mesures qui ont été effectuées au laboratoire IMS nous ont permis d'effectuer cette normalisation.

A titre d'exemple nous présentons sur la figure 19 les résultats normalisés de la radio, du téléphone portable à 900 MHz et 1800MHz.

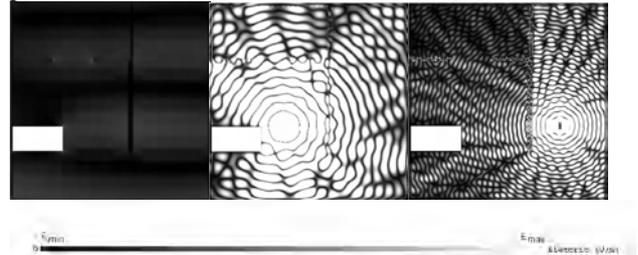

*Figure 19 : Comparaison des niveaux de champs électriques émis par la radio, du téléphone portable à 900 MHz et 1800MHz.*

## 7 SUPERPOSITION DES CONTRIBUTIONS DES DIFFERENTES APPLICATIONS

Nous avons présenté des résultats issus de nos modèles HFSS et Microstripes. A l'aide de ces outils de simulation, il est impossible de calculer puis de tracer simultanément des répartitions de champs électromagnétiques de fréquences différentes. La solution retenue a été d'obtenir ces superpositions en traitant après calculs les cartographies de champs électromagnétiques en tant qu'images. Nous avons développé une démarche pour traiter les images des cartographies de champ électromagnétique à l'aide du logiciel Aphelion [5]. Ces répartitions de champ ainsi que leur superposition ont été associés à une interface graphique. Cette dernière permet de choisir les différentes applications que nous souhaitons visualiser.

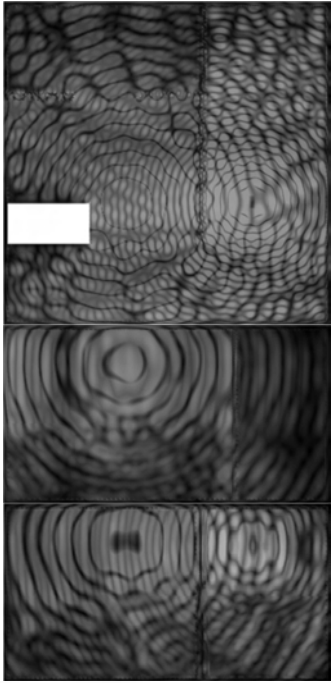

*Figure 20 : Superposition des champs électriques des différentes applications suivant trois plans de coupe.*

## 8   CONCLUSION

Les simulations numériques que nous avons effectuées sont des supports de cours très intéressants pour les étudiants. L'outil pédagogique complet que nous avons développé (superposition des champs, géométries 3D, interface graphique) a de nombreuses applications. Tout d'abord en cours, il permettra de sensibiliser les étudiants à notre environnement électromagnétique et à la notion de compatibilité électromagnétique. Cet outil et les résultats des études menées par de nombreuses équipes (dont l'équipe bioélectromagnétisme du département MCM du laboratoire IMS) leur permettront aussi de vérifier l'innocuité des applications des réseaux sans fil. Toutefois, selon le principe de précaution, des recommandations existent quant à l'utilisation de certaines de ces applications (téléphonie mobile) [6][7][8][9]. Cet outil pourrait aussi être associé dans le cadre de travaux pratiques à des mesures à l'aide d'un analyseur de spectre et à une série d'antennes. Enfin, il sera utilisé lors d'évènements ouverts au public tels que la fête de la science.